\documentclass[a4paper,11pt]{article}
\pdfoutput=1 
\usepackage{jheppub} 

\usepackage[T1]{fontenc} 
\usepackage[utf8]{inputenc} 
\usepackage{microtype} 
\usepackage{mdframed} 
\usepackage{graphicx}
\usepackage{todonotes} 

\usepackage{amsmath}
\usepackage{amsfonts}
\usepackage{amssymb}
\usepackage{stmaryrd}
\usepackage{mathtools}
\usepackage{mathrsfs}
\usepackage{scalerel}
\usepackage{eucal} 
\usepackage{mleftright} \mleftright
\usepackage{tensor} 
\usepackage{braket} 
\usepackage{mathrsfs} 
\usepackage{bbold}
\usepackage{color}
\usepackage{braket}

\newcommand{\comment}[1]{}

\providecommand*{\dd}{\mathop{}\!\mathrm{d}}
\renewcommand*{\dd}{\mathop{}\!\mathrm{d}}

\providecommand*{\R}{{\mathbb{R}}}
\renewcommand*{\R}{{\mathbb{R}}}

\providecommand{\Lt}{{\tt L}}
\renewcommand{\Lt}{{\tt L}}
\providecommand{\Mt}{{\tt M}}
\renewcommand{\Mt}{{\tt M}}
\providecommand{\Ut}{{\tt U}}
\renewcommand{\Ut}{{\tt U}}
\providecommand{\Vt}{{\tt V}}
\renewcommand{\Vt}{{\tt V}}

\providecommand{\Et}{{\tt E}}
\renewcommand{\Et}{{\tt E}}

\newcommand{\chemM}{\mu_{\textrm{\tiny M}}}
\newcommand{\chemV}{\mu_{\textrm{\tiny V}}}
\newcommand{\chemL}{\mu_{\textrm{\tiny L}}}
\newcommand{\chemU}{\mu_{\textrm{\tiny U}}}
\newcommand{\chemP}{\mu_\mathcal{P}}
\newcommand{\chemJ}{\mu_\mathcal{J}}

\DeclareMathOperator{\extdm}{d}
\newcommand{\extd}{\extdm \!}
\DeclareRobustCommand{\rchi}{{\mathpalette\irchi\relax}}
\newcommand{\irchi}[2]{\raisebox{\depth}{$#1\chi$}}

\newcommand{\eq}[2]{\begin{equation} #1 \label{#2} \end{equation}}

\title{Higher-Spin Flat Space Cosmologies with Soft Hair}

\author[\spadesuit]{Martin Ammon,}
\author[\heartsuit\diamondsuit]{Daniel Grumiller,}
\author[\heartsuit]{Stefan Prohazka,}
\author[\clubsuit]{Max Riegler,}
\author[\heartsuit]{and Raphaela Wutte}

\affiliation[\spadesuit]{Theoretisch-Physikalisches Institut, Friedrich-Schiller University of Jena, Max-Wien-Platz 1, D-07743 Jena, Germany, Europe}
\affiliation[\heartsuit]{Institute for Theoretical Physics, TU Wien, Wiedner Hauptstrasse 8--10/136, A-1040 Vienna, Austria, Europe}
\affiliation[\diamondsuit]{CMCC-Universidade Federal do ABC, Santo Andr\'e, S.P. Brazil}
\affiliation[\clubsuit]{Universit{\'e} libre de Bruxelles, Boulevard du Triomphe (Campus de la Plaine), 1050 Bruxelles, Belgium, Europe}

\emailAdd{martin.ammon@uni-jena.de}
\emailAdd{grumil@hep.itp.tuwien.ac.at}
\emailAdd{prohazka@hep.itp.tuwien.ac.at}
\emailAdd{Max.Riegler@ulb.ac.be}
\emailAdd{rwutte@hep.itp.tuwien.ac.at}

\preprint{TUW--17--01}

\abstract{
We present and discuss near horizon boundary conditions for flat space higher-spin gravity in three dimensions. As in related work our boundary conditions ensure regularity of the solutions independently of the charges. The asymptotic symmetry algebra is given by a set of $\hat{\mathfrak{u}}(1)$ current algebras. The associated charges generate higher-spin soft hair. We derive the entropy for solutions that are continuously connected to flat space cosmologies and find the same result as in the spin-2 case: the entropy is linear in the spin-2 zero-mode charges and independent from the spin-3 charges. Using twisted Sugawara-like constructions of higher-spin currents we show that our simple result for entropy of higher-spin flat space cosmologies coincides precisely with the complicated earlier results expressed in terms of higher-spin zero mode charges.
}

\begin{document}
\maketitle
\flushbottom

\section{Introduction}
\label{sec:introduction}
Higher-spin theories provide useful insights into aspects of the holographic principle \cite{Klebanov:2002ja, Sezgin:2002rt, Giombi:2012ms}. Particularly three-dimensional higher-spin theories are useful in this context, since they can be formulated as Chern--Simons theories \cite{Blencowe:1988gj} with specific boundary conditions \cite{Henneaux:2010xg,Campoleoni:2010zq,Gaberdiel:2011wb,Campoleoni:2011hg}. Developments in three-dimensional higher-spin theories include the discovery of minimal model holography \cite{Gaberdiel:2010pz, Gaberdiel:2012uj}, higher-spin black holes \cite{Gutperle:2011kf, Castro:2011fm, Ammon:2012wc,Bunster:2014mua}, non-AdS holography \cite{Gary:2012ms,Gutperle:2013oxa,Bergshoeff:2016soe}, higher-spin holographic entanglement entropy \cite{Ammon:2013hba, deBoer:2013vca} and particularly flat space higher-spin theories \cite{Afshar:2013vka, Gonzalez:2013oaa}, the main topic of the present work.

An interesting and potentially confusing aspect of higher-spin theories is that the metric and associated notions like curvature singularities or horizons are not gauge-invariant entities. Nevertheless, there are field configurations that most naturally are interpreted as (higher-spin) black holes or (higher-spin) cosmologies, i.e., as solutions with some characteristic temperature and entropy. Many of the physical questions inspired by black holes and cosmologies addressed in spin-2 gravity can also be addressed in a higher-spin context, sometimes in a straightforward way, but quite often with surprising generalizations and qualitatively new features emerging from the massless higher-spin interactions. In the present work we focus on one particular issue, namely on ``soft hair'' in flat space higher-spin theories in three dimensions. 

The notion of ``soft hair'' was introduced in a spin-2 context in \cite{Hawking:2016msc} and refers to zero energy excitations on black hole horizons. To explicitly construct soft hair excitations, but more generally to address any question that requires the existence of a black hole as part of the question, it is then useful to have boundary conditions that ensure regular horizons for all configurations. While these boundary conditions can be re-interpreted as asymptotic fall-off conditions of Brown--Henneaux type \cite{Brown:1986nw}, they take their most natural form if expanded around the horizon. Thus, we shall refer to them as ``near horizon boundary conditions''. 

In AdS$_3$ different near horizon boundary conditions were proposed independently in \cite{Donnay:2015abr}, \cite{Afshar:2015wjm} and \cite{Afshar:2016wfy}. In this work we focus on the latter approach, since it leads to the simplest symmetry algebras and due to the Chern--Simons formulation used in \cite{Afshar:2016wfy} it is most suitable for generalizations to higher-spins in AdS$_3$ \cite{Grumiller:2016kcp} or flat space \cite{Afshar:2016kjj}. The main goal of the present work is to further generalize these results to higher-spins in flat space.

Our main results are new boundary conditions suitable for constructing soft higher-spin hair on flat space cosmologies and a remarkably simple expression for their entropy,
\eq{
S = 2\pi \big(J_0^+ + J_0^-\big)
}{eq:1}
where $J^\pm_0$ are the spin-2 zero mode charges. Precisely the same result was found in AdS$_3$ Einstein gravity \cite{Afshar:2016wfy}, in higher derivative gravity \cite{Setare:2016vhy}, in AdS$_3$ higher-spin gravity \cite{Grumiller:2016kcp} and in flat space Einstein gravity \cite{Afshar:2016kjj}, where the soft hair on black hole horizons is replaced by soft hair on flat space cosmological horizons. The simplicity and universality of the result for the entropy \eqref{eq:1} is intriguing.

This paper is organized as follows.
In section \ref{nearhorizon} we present our near horizon boundary conditions and the associated symmetries.
In section \ref{mapfromdiagonaltohighest} we provide a map from diagonal to highest weight gauge.
In section \ref{blackholeentropy} we calculate the entropy of higher-spin flat space cosmologies and exploit the map from the previous section to match our simple result for entropy with the complicated results appearing in the literature. 
In section \ref{se:metric} we translate from Chern--Simons into second order formulation and give explicit results for metric and spin-3 field.
Before we conclude we discuss in section \ref{sec:extens-high-spin} the generalization to fields with spin greater than $3$.
The appendices provide details on $\mathfrak{isl}(N,\mathbb{R})$ and $\mathfrak{ihs}[\lambda]$ algebras.

\section{Near horizon boundary conditions and symmetries}
\label{nearhorizon}
Asymptotically flat higher-spin gravity in three spacetime dimensions is conveniently formulated in terms of a Chern--Simons theory. Restricting for simplicity to spin-3 gravity, the action reads
\begin{equation}
    I[\mathcal{A}] = \frac{k}{4\pi} \int \langle \textrm{CS}(\mathcal{A})\rangle \, , \qquad \textrm{with} \quad  \textrm{CS}(\mathcal{A}) = \mathcal{A} \wedge \extd \mathcal{A} + \frac{2}{3} \mathcal{A}\wedge \mathcal{A} \wedge \mathcal{A} \, ,
\end{equation}
with Chern-Simons coupling $k= 1/(4G_N)$ and gauge field $\mathcal{A}$ valued in $\mathfrak{isl}(3,\mathbb{R})$. The generators of $\mathfrak{isl}(3,\mathbb{R})$ are denoted by ${\Lt}_i, {\Mt}_i, {\Ut}_m, {\Vt}_m$ 
with $i \in \{ -1, 0, 1 \}$ and $m \in \{ -2, -1, 0, 1, 2\}$. While ${\Lt}_i$ and ${\Mt}_i$ generate Lorentz-Transformations and translations, respectively, ${\Ut}_m$ and ${\Vt}_m$ generate associated spin-3 transformations. We refer the reader to appendix \ref{append} for the commutation relations satisfied by the generators as well as for the definition of the non-degenerate invariant symmetric bilinear form $\langle \dots \rangle$. Moreover, we use coordinates $(r, v, \varphi)$, where $r$ denotes the radial coordinate, $v$ the advanced time and $\varphi$ the angular coordinate.

In order to specify our boundary conditions we first use some of the gauge freedom at our disposal to fix the radial dependence of the connection $\mathcal{A}$ as
    \begin{equation}
\label{eq:gaugeans}
        \mathcal{A}=b^{-1}(a+\extd \,) \, b\,,
    \end{equation}
where the radial dependence is encoded in the group element $b$ as~\cite{Afshar:2016kjj}
    \begin{equation}
\label{eq:gaugeb}
        b=\exp\left( \frac{1}{\chemP} \, \Mt_1\right) \, \exp\left( \frac{r}{2} \, \Mt_{-1}\right) \,.
    \end{equation}
and the connection $a$ reads
\begin{equation}
a = a_v  \extd v + a_\varphi \extd \varphi \, .
\end{equation}
We propose the following new near-horizon boundary conditions\footnote{%
The relation to the notation used in \cite{Afshar:2016kjj} is given by $a=-\chemP$, $\omega=\mathcal{J}$, $\Omega=\chemJ$, $\gamma=\mathcal{P}$.
}
    \begin{subequations}\label{eq:FSCSpin3BCs}
    \begin{align}
        a_\varphi&=\mathcal{J} \, \Lt_0+\mathcal{P} \, \Mt_0+\mathcal{J}^{(3)} \, \Ut_0+\mathcal{P}^{(3)} \, \Vt_0 \, ,  \\
        a_v&=\mu_\mathcal{P} \, \Lt_0+\mu_\mathcal{J} \, \Mt_0+\mu_\mathcal{P}^{(3)} \, \Ut_0+\mu_\mathcal{J}^{(3)} \, \Vt_0\,.
    \end{align}
    \end{subequations}
All the functions appearing in \eqref{eq:FSCSpin3BCs} are in principle arbitrary functions of the advanced time $v$ and the angular coordinate $\varphi$. The functions $\mu_a$ are identified as chemical potentials and thus are fixed in such a way that $\delta\mu_a=0$. The equations of motion
\begin{equation}
F=\extd\mathcal{A}+[\mathcal{A},\mathcal{A}]=0
\end{equation}
put further constraints on the functions $\mathcal{J},\mathcal{P}$ as well as $\mathcal{J}^{(3)},\mathcal{P}^{(3)}$ that can be interpreted as holographic Ward identities. These constraints force the state dependent functions to obey the following time evolution equations
    \begin{equation}\label{eq:EOM}
        \partial_v\mathcal{J}=\partial_\varphi\mu_\mathcal{P},\quad         \partial_v\mathcal{P}=\partial_\varphi\mu_\mathcal{J},\quad        \partial_v\mathcal{J}^{(3)}=\partial_\varphi\mu_\mathcal{P}^{(3)},\quad         \partial_v\mathcal{P}^{(3)}=\partial_\varphi\mu_\mathcal{J}^{(3)}\,.
    \end{equation}
In particular, for $\varphi$-independent chemical potentials the holographic Ward identities \eqref{eq:EOM} imply conservation of all the state dependent functions.

\subsection{Canonical charges and near horizon symmetry algebra}
The next step in the asymptotic symmetry analysis is to determine the gauge transformations $\delta_\epsilon\mathcal{A}=\extd\epsilon+[\mathcal{A},\epsilon]$
that preserve the boundary conditions \eqref{eq:gaugeans}--\eqref{eq:FSCSpin3BCs}. The gauge parameters $\epsilon$ that encode such transformations are given by
    \begin{equation}\label{eq:BCPGTS}
        \epsilon=b^{-1}(\epsilon_\mathcal{P} \Lt_0+\epsilon_\mathcal{J} \Mt_0+\epsilon_\mathcal{P}^{(3)} \Ut_0+\epsilon_\mathcal{J}^{(3)} \Vt_0) \,b\,.
    \end{equation}
As a consequence also the infinitesimal transformation behavior of the state dependent functions takes a particularly simple form
    \begin{equation}       \delta\mathcal{J}=\partial_\varphi\epsilon_\mathcal{P},\quad\delta\mathcal{P}=\partial_\varphi\epsilon_\mathcal{J},\quad\delta\mathcal{J}^{(3)}=\partial_\varphi\epsilon_\mathcal{P}^{(3)},\quad\delta\mathcal{P}^{(3)}=\partial_\varphi\epsilon_\mathcal{J}^{(3)}\,.
    \end{equation}  
Moreover, the conserved charges $Q  \left[\epsilon \right]$ associated to boundary conditions preserving transformations may be computed via
the Regge-Teitelboim approach \cite{Regge:1974zd}, where their variation is given by
\begin{equation}
\label{eq: generalQ}
\delta Q \left[\epsilon \right]= \frac{k}{2\pi} \int \extd \varphi \braket{\epsilon\, \delta \mathcal{A}_{\varphi}}.
\end{equation}
Evaluating this expression for our case yields 
    \begin{equation}
        \delta Q[\epsilon]=\frac{k}{2\pi}\int\extd\varphi \, \langle\epsilon \, \delta \mathcal{A}_\varphi\rangle=
\frac{k}{2\pi}\int\extd\varphi\left(
\epsilon_\mathcal{J} \delta\mathcal{J} + \epsilon_\mathcal{P} \delta\mathcal{P} + \frac{4}{3} \epsilon_\mathcal{J}^{(3)} \delta\mathcal{J}^{(3)}+\frac{4}{3} \epsilon_\mathcal{P}^{(3)} \delta\mathcal{P}^{(3)} 
\right)\,. \label{eq: deltaQ}
    \end{equation}
The global charges may now be obtained by functionally integrating \eqref{eq: deltaQ},
    \begin{equation}
        Q[\epsilon]=\frac{k}{2\pi}\int\extd\varphi \, \langle\epsilon\, \mathcal{A}_\varphi\rangle = \frac{k}{2\pi}\int\extd\varphi \left( 
 \epsilon_\mathcal{J} \mathcal{J}+ \epsilon_\mathcal{P} \mathcal{P}+\frac{4}{3} \epsilon_\mathcal{J}^{(3)} \mathcal{J}^{(3)}+\frac{4}{3} \epsilon_\mathcal{P}^{(3)} \mathcal{P}^{(3)}
\right) \, .
    \end{equation}
After having determined the canonical boundary charges, their Dirac bracket algebra can be read off from their infinitesimal transformation behavior using
    \begin{equation}
        \delta_{Y}Q\left[X\right]=\{Q\left[X\right],Q\left[Y\right]\} \, .
    \end{equation}
This yields
    \begin{equation}
        \{\mathcal{J}(\varphi),\mathcal{P}(\bar{\varphi})\}=\frac{k}{2\pi}\partial_\varphi\delta(\varphi-\bar{\varphi})\,,\qquad \quad \{\mathcal{J}^{(3)}(\varphi),\mathcal{P}^{(3)}(\bar{\varphi})\}=\frac{2k}{3\pi}\partial_\varphi\delta(\varphi-\bar{\varphi})\,,
    \end{equation}
where all other Dirac brackets vanish. Expanding into Fourier modes 
    \begin{subequations}\label{eq:CurrentFourierModes}
	\begin{align}
		\mathcal{J}(\varphi)&=\frac{1}{k}\sum\limits_{n\in\mathbb{Z}}J_ne^{-i n \varphi}&
		\mathcal{P}(\varphi)&=\frac{1}{k}\sum\limits_{n\in\mathbb{Z}}P_ne^{-i n \varphi}\\
		\mathcal{J}^{(3)}(\varphi)&=\frac{3}{4k}\sum\limits_{n\in\mathbb{Z}}J_n^{(3)}e^{-i n \varphi}&
		\mathcal{P}^{(3)}(\varphi)&=\frac{3}{4k}\sum\limits_{n\in\mathbb{Z}}P_n^{(3)}e^{-i n \varphi}
	\end{align}
	\end{subequations}
(with the usual decomposition of the $\delta$-function, $2\pi \delta(\varphi-\bar{\varphi})=\sum e^{-in(\varphi-\bar{\varphi})}$),
and replacing the Dirac brackets by commutators using $i\{\cdot,\cdot\}\rightarrow [\cdot,\cdot]$ we obtain the following asymptotic symmetry algebra for the boundary conditions \eqref{eq:gaugeans}--\eqref{eq:FSCSpin3BCs}
    \begin{equation}\label{eq:FourierASA}
        [J_n,P_m]=k \, n \, \delta_{n+m,0}\,,\qquad\quad [J_n^{(3)},P_m^{(3)}]=\frac{4k}{3} \, n \, \delta_{n+m,0}\,
    \end{equation}
with all other commutators vanishing. At this point it should also be noted that the algebra \eqref{eq:FourierASA} can be brought to the same form as in \cite{Grumiller:2016kcp} by making the redefinitions
    \begin{equation}
        J_{\pm n}^\pm=\frac{1}{2}(P_n\pm J_n)\,,\qquad J_{\pm n}^{(3)\pm}=\frac{1}{2}(P_n^{(3)}\pm J_n^{(3)})\,.
    \end{equation}
The generators $J_n^\pm$ and $J_n^{{(3)\pm}}$ then satisfy
    \begin{subequations}
    \begin{align}
        [J^+_n,J^+_m]&=[J^-_n,J^-_m]=\frac{k}{2}n\delta_{n+m,0}\,,\qquad\quad &[J^+_n,J^-_m]=0 \, ,\\
        [J^{(3)+}_n,J^{(3)+}_m]&=[J^{(3)-}_n,J^{(3)-}_m]=\frac{2k}{3}n\delta_{n+m,0} \, , &[J^{(3)+}_n,J^{(3)-}_m]=0 \, .  
    \end{align}
   \end{subequations}
In particular, we obtain in total four $\hat{\mathfrak{u}}(1)$ current algebras, two of which have level $k/2$ and the remaining two have level $2k/3.$

\subsection{Soft Hair}\label{sec:SoftHair}

In this subsection we show that the states generated by acting with arbitrary combinations of near horizon symmetry generators \eqref{eq:FourierASA} on some reference state all have the same energy and thus correspond to soft hair excitations of that reference state. In order to show this we first determine the Hamiltonian in terms of near horizon variables, then proceed in building modules using \eqref{eq:FourierASA}, and finally show that all states in these modules have the same energy eigenvalue.

The Hamiltonian is associated to the charge that generates time translations. In the metric formulation this would correspond to the Killing vector $\partial_v$. Since the gauge transformations \eqref{eq:BCPGTS} are related on-shell to the asymptotic Killing vectors $\xi^\mu$ via $\epsilon=\xi^\mu \mathcal{A}_\mu$, the variation of the charge associated to translations in the advanced time coordinate $v$ can be determined via
    \begin{align}\label{eq:HamiltonianVar}
        \delta H:=\delta Q[\partial_v]&=\frac{k}{2\pi}\int\extd\varphi \, \langle\xi^v\mathcal{A}_v \, \delta\mathcal{A}_\varphi\rangle=\frac{k}{2\pi}\int\extd\varphi \, \langle\mathcal{A}_v \, \delta\mathcal{A}_\varphi\rangle\nonumber\\
        &=\frac{k}{2\pi}\int\extd\varphi\left(\mu_\mathcal{J}\delta\mathcal{J}+\mu_\mathcal{P}\delta\mathcal{P}+\frac{4}{3}\mu_\mathcal{J}^{(3)}\delta\mathcal{J}^{(3)}+\frac{4}{3}\mu_\mathcal{P}^{(3)}\delta\mathcal{P}^{(3)}\right)\,.
    \end{align}
This expression can be trivially functionally integrated to yield the Hamiltonian
    \begin{equation}
        H=\frac{k}{2\pi}\int\extd\varphi\left(\mu_\mathcal{J}\mathcal{J}+\mu_\mathcal{P}\mathcal{P}+\frac{4}{3}\mu_\mathcal{J}^{(3)}\mathcal{J}^{(3)}+\frac{4}{3}\mu_\mathcal{P}^{(3)}\mathcal{P}^{(3)}\right)\,.
    \end{equation}
For constant chemical potentials $\mu_a$ and $\mu_a^{(3)}$ the Hamiltonian reduces to
    \begin{equation}\label{eq:Hamiltonian}
        H=\left(\mu_\mathcal{J}J_0+\mu_\mathcal{P}P_0+\frac{4}{3}\mu_\mathcal{J}^{(3)}J^{(3)}_0+\frac{4}{3}\mu_\mathcal{P}^{(3)}P^{(3)}_0\right)\,.
    \end{equation}
After having determined the Hamiltonian the next step in our analysis is to build modules using \eqref{eq:FourierASA}. There are two ways of building modules relevant to our analysis. One is via \emph{highest weight} representations wheres the other one uses a construction similar to  \emph{induced} representations.\\
We first start with modules built from \emph{highest weight} representations of \eqref{eq:FourierASA}. Assume that there is a highest weight (vacuum) state $|0\rangle$ satisfying
    \begin{equation}
        J_n|0\rangle=P_n|0\rangle=J_n^{(3)}|0\rangle=P^{(3)}_n|0\rangle=0,\qquad\forall\,n\geq0\,.
    \end{equation}
New states can then be constructed from such a vacuum state by repeated application of operators with negative Fourier mode number as
    \begin{equation}\label{eq:HighestWeightStates}
        |\psi(\{p\})\rangle\sim\prod\limits_{n_i>0}J_{-n_i}\prod\limits_{n_i^{(3)}>0}J_{-n_i^{(3)}}^{(3)}\prod\limits_{m_i>0}P_{-m_i}\prod\limits_{m_i^{(3)}>0}P_{-m_i^{(3)}}^{(3)}|0\rangle\,,
    \end{equation}
where $\{p\}\equiv\{n_i,n_i^{(3)},m_i,m_i^{(3)}\}$. Since the Hamiltonian is a linear combination of $J_0$, $P_0$, $J^{(3)}_0$ and $P_0^{(3)}$, it is evident that the Hamiltonian commutes with any element appearing in the asymptotic symmetry algebra \eqref{eq:FourierASA}. Thus when acting with $H$ on any $\psi(\{p\})$ one obtains the same value for the energy for all possible $\{p\}$'s. This proves our claim that the states $|\psi(\{p\})\rangle$ are ``soft hair'' of the vacuum; similar considerations apply when replacing the vacuum $|0\rangle$ with any other state, such as some flat space cosmology, which can then be decorated with soft spin-2 and spin-3 hair.
\\
Now we investigate the same issue for modules built from representations that are similar in spirit to the \emph{induced} representations found in flat space holography (see e.g.~\cite{Campoleoni:2015qrh,Campoleoni:2016vsh}). 
In the following we consider all ``boosted'' states that can be built from a ``rest frame'' state $|\Omega\rangle$ via
    \begin{equation}\label{eq:FSCStates}
        |\psi(\{q\})\rangle\sim\prod\limits_{n_i}\left(J_{n_i}\right)\prod\limits_{n_i^{(3)}}\left(J^{(3)}_{n_i^{(3)}}\right)|\Omega\rangle,
    \end{equation}
where $\{q\}\equiv\{n_i,n_i^{(3)}\}$. 
For a given ``rest frame'' state $|\Omega\rangle$ one can generate [$\hat{\mathfrak{u}}(1)$] ``boosted'' states as written in \eqref{eq:FSCStates}. In addition this ``rest frame'' state has to satisfy
    \begin{equation}
        P_n|\Omega\rangle=P_n^{(3)}|\Omega\rangle=0,\quad\forall\,n\in\mathbb{Z}\,.
    \end{equation}
One way to argue such representations is via taking an ultra-relativistic limit of the highest-weight representations used in \cite{Grumiller:2016kcp}. On the level of generators the ASA\footnote{We focus only on the generators $J_n$ and $P_n$. The argument can be repeated in the exact same way for $J_n^{(3)}$ and $P_n^{(3)}$.} in \cite{Grumiller:2016kcp} and the one in \eqref{eq:FourierASA} are related via an ultra-relativistic boost that can be incorporated as
    \begin{equation}
        J^\pm_{\pm n}=\frac{1}{2}\left(\frac{P_n}{\epsilon}\pm J_n\right)\,,
    \end{equation}
in the limit $\epsilon\rightarrow0$. By looking at highest-weight representations built from $J^\pm_n$ one finds that in terms of the generators $P_n$ and $J_n$ one has
    \begin{equation}\label{eq:URHWLimit}
        J^\pm_n|\Omega\rangle=\frac{1}{2}\left(\frac{P_{\pm n}}{\epsilon}\pm J_{\pm n}\right)|\Omega\rangle=0,\quad\forall\,n\geq0\,.
    \end{equation}
In order to satisfy these relations when $\epsilon\rightarrow0$ one finds that, indeed, acting with $P_n$ on $|\Omega\rangle$ has to be zero for all values of $n$, whereas one can act with $J_n$ on $|\Omega\rangle$ without spoiling \eqref{eq:URHWLimit}.
One can now again act with the Hamiltonian \eqref{eq:Hamiltonian} on all states in the module \eqref{eq:FSCStates} and using the same line of argument as for the highest weight representations one finds again that all states have the same energy eigenvalue and can thus be interpreted as soft excitations as well.

Thus, the soft hair property does not depend on whether highest weight or representations of the form \eqref{eq:FSCStates} are used. Moreover, since the Hamiltonian is an element of the Cartan subalgebra of $\mathfrak{isl}(3,\mathbb{R})$, one can even conclude that the soft hair property is independent of \emph{any} representation that can be built via acting on some reference state using the near horizon symmetry generators.

\section{Relating near horizon and asymptotic symmetries}
\label{mapfromdiagonaltohighest}

In order to show that the spin-2 and spin-3 charges of higher-spin cosmological solutions in flat space emerge as composite operators constructed from the $\hat{\mathfrak{u}}(1)$ ones, we have to relate the boundary conditions we presented in this work \eqref{eq:FSCSpin3BCs} with the boundary conditions that describe a flat space cosmology with spin-2 and spin-3 hair. Thus it is first necessary to describe both boundary conditions by the same set of variables.

Flat space cosmologies with spin-2 and spin-3 hair including chemical potentials are given by the following connection \cite{Gary:2014ppa}
    \begin{equation}
        \tilde{\mathcal{A}}=\tilde{b}^{-1}(\tilde{a}+\extd \, )\tilde{b}\,,
    \end{equation}
with $\tilde{b}=\exp(\frac{r}{2} \Mt_{-1})$ and
	\begin{subequations}\label{eq:FSCSpin3Connection}
	\begin{align}
		\tilde{a}_\varphi &= \Lt_1-\frac{\mathcal{M}}{4} \Lt_{-1}-\frac{\mathcal{N}}{2} \Mt_{-1}+\frac{\mathcal{V}}{2} \Ut_{-2}+\mathcal{Z} \Vt_{-2}\\
		\tilde{a}_v &= a_v^{(0)} + a_v^{(\chemM)}  + a_v^{(\chemL)}  + a_v^{(\chemV)}  + a_v^{(\chemU)}
	\end{align}
	\end{subequations}
where
	\begin{subequations}\label{eq:FSCSpin3ConectionDetail}
	\begin{align}
          a_v^{(0)} &= \Mt_1-\frac{\mathcal{M}}{4} \Mt_{-1}+\frac{\mathcal{V}}{2} \Vt_{-2}\\
          a_v^{(\chemM)} &= \chemM\, \Mt_1 -\chemM^\prime\, \Mt_0 + \tfrac12\,\big(\chemM'' - \tfrac12{\cal M} \chemM \big)\, \Mt_{-1} + \tfrac 1 2 \,{\cal V}\, \chemM\, \Vt_{-2} \\
          a_v^{(\chemL)} &= a_v^{(\chemM)}\big|_{M\to L}  - \tfrac 1 2 \, {\cal N}\,\chemL\, \Mt_{-1} + {\cal Z}\,\chemL\, \Vt_{-2}\displaybreak[1]\\
          a_v^{(\chemV)} &= \chemV \, \Vt_2 - \chemV^\prime\, \Vt_1   + \tfrac12 \,\big(\chemV'' - {\cal M} \chemV\big)\, \Vt_0 + \tfrac16\,\big(- \chemV'''+ {\cal M}^\prime \chemV  + \tfrac{5}{2} {\cal M} \chemV^\prime\big)\, \Vt_{-1}   \nonumber \\
&\quad  + \tfrac{1}{24}\,\big(\chemV''''  - 4 {\cal M} \chemV'' - \tfrac72{\cal M}^\prime \chemV^\prime  + \tfrac32 {\cal M}^2 \chemV - {\cal M}'' \chemV \big)\,\Vt_{-2} - 4{\cal V}\, \chemV\, \Mt_{-1} \\
          a_v^{(\chemU)} &= a_v^{(\chemV)}\big|_{M\to L}  - 8{\cal Z}\,\chemU\, \Mt_{-1}  - {\cal N}\,\chemU\, \Vt_0 + \big(\tfrac56 {\cal N} \chemU^\prime+\tfrac13{\cal N}^\prime \chemU \big)\, \Vt_{-1}  \nonumber \\
& \quad + \big(- \tfrac13{\cal N} \chemU'' - \tfrac{7}{24} {\cal N}^\prime \chemU^\prime   - \tfrac{1}{12}{\cal N}'' \chemU + \tfrac14 {\cal M} {\cal N} \chemU \big)\, \Vt_{-2}\,. 
	\end{align}
	All functions appearing in \eqref{eq:FSCSpin3Connection} and \eqref{eq:FSCSpin3ConectionDetail} are free functions of $v$ and $\varphi$. As such a prime denotes a derivative with respect to $\varphi$ and a dot a derivative with respect to $v$. The subscript $M \to L$ denotes that in the corresponding quantity all generators and chemical potentials are replaced as ${\Mt_n\to \Lt_n}$, $\Vt_n\to \Ut_n$, $\chemM\to\chemL$ and $\chemV\to\chemU$, i.e.
	\begin{align}
          a_v^{(\chemM)}\big|_{M\to L} &= \chemL\, \Lt_1 -\chemL^\prime\, \Lt_0 + \tfrac12\,\big(\chemL''-\tfrac12{\cal M} \chemL \big) \,\Lt_{-1} + \tfrac12\,{\cal V} \,\chemL\, \Ut_{-2} \\
          a_v^{(\chemV)}\big|_{M\to L} &=  \chemU\, \Ut_2 - \chemU^\prime\, \Ut_1   + \tfrac12 \,\big(\chemU'' - {\cal M} \chemU\big)\, \Ut_0 + \tfrac16\,\big(- \chemU'''+ {\cal M}^\prime \chemU  + \tfrac{5}{2} {\cal M} \chemU^\prime\big)\, \Ut_{-1}   \nonumber \\
&\quad  + \tfrac{1}{24}\,\big(\chemU''''  - 4 {\cal M} \chemU'' - \tfrac72{\cal M}^\prime \chemU^\prime  + \tfrac32 {\cal M}^2 \chemU - {\cal M}'' \chemU \big)\,\Ut_{-2} - 4{\cal V} \,\chemU\, \Lt_{-1}\,.
    \end{align}
    \end{subequations}
The next step is to find an appropriate gauge transformation that maps the connection $a$ in \eqref{eq:FSCSpin3BCs} to the connection $\tilde{a}$ in \eqref{eq:FSCSpin3Connection} via $\tilde{a}=g^{-1}(a+\extd \,)g$.
After a fair amount of algebraic manipulation one can find the following group element that provides the appropriate map as $g=g^{(1)}g^{(2)}$ with
    \begin{subequations}
    \begin{align}
        g^{(1)} &= \exp\left[\mathfrak{l} \, \Lt_1+\mathfrak{m} \, \Mt_1+\mathfrak{u}_1 \,  \Ut_1+\mathfrak{v}_1 \, \Vt_1+\mathfrak{u}_2 \, \Ut_2+\mathfrak{v}_2 \, \Vt_2\right]\\
        g^{(2)} &= \exp\left[-\frac{\mathcal{J}}{2} \, \Lt_{-1}-\frac{\mathcal{J}^{(3)}}{3} \, \Ut_{-1}+\frac{1}{6}\left(\mathcal{J}\mathcal{J}^{(3)}+\frac{\mathcal{J}^{(3)'}}{2}\right) \, \Ut_2\right.\\            &\qquad\quad\,\,\left.-\frac{\mathcal{J}}{2} \, \Mt_{-1}-\frac{\mathcal{P}^{(3)}}{3} \, \Vt_{-1}+\frac{1}{6}\left(\mathcal{P}\mathcal{J}^{(3)}+\mathcal{J}\mathcal{P}^{(3)}+\frac{\mathcal{P}^{(3)'}}{2}\right) \, \Vt_{-2}\right]\,.
    \end{align}
    \end{subequations}
The functions $\mathfrak{l}$, $\mathfrak{m}$, $\mathfrak{u}_a$ and $\mathfrak{v}_a$ depend on $v$ and $\varphi$ only and have to satisfy
    \begin{subequations}
    \begin{align}
        \mathfrak{l}' &= 1+ \mathfrak{l}\mathcal{J}+2\mathfrak{u}_1\mathcal{J}^{(3)}\\
        \mathfrak{m}' &= \mathfrak{l}\mathcal{P}+\mathfrak{m}\mathcal{P}+2\mathfrak{u}_1\mathcal{P}^{(3)}+2\mathfrak{v}_1\mathcal{J}^{(3)}\\
        \mathfrak{u}_1' &= \mathfrak{u}_1\mathcal{J}+2\mathfrak{l}\mathcal{J}^{(3)}\\
        \mathfrak{v}_1' &= \mathfrak{u}_1\mathcal{P}+\mathfrak{v}_1\mathcal{J}+2\mathfrak{l}\mathcal{P}^{(3)}+2\mathfrak{m}\mathcal{J}^{(3)}\\
        \mathfrak{u}_2' &= -\frac{\mathfrak{u}_1}{2}+2\mathfrak{u}_2\mathcal{J}\\
        \mathfrak{v}_2' &= -\frac{\mathfrak{v}_1}{2}+2\mathfrak{u}_2\mathcal{P}+2\mathfrak{v}_2\mathcal{J}\,,        
    \end{align}
    \end{subequations}
and 
    \begin{subequations}
    \begin{align}
        \chemL &= \frac{4}{3}\chemU\mathcal{J}^{(3)}-\mu_\mathcal{P}\mathfrak{l}-2\mu_\mathcal{P}^{(3)}\mathfrak{u}_1+\dot{\mathfrak{l}}\\
        \chemM &= \frac{4}{3}\chemU\mathcal{P}^{(3)}+\frac{4}{3}\chemV\mathcal{J}^{(3)}-\mu_\mathcal{P}\mathfrak{m}-\mu_\mathcal{J}\mathfrak{l}-2\mu_\mathcal{P}^{(3)}\mathfrak{v}_1-2\mu_\mathcal{J}^{(3)}\mathfrak{u}_1+\dot{\mathfrak{m}}\displaybreak[1]\\
        \chemU &= -2\mu_\mathcal{P}\mathfrak{u}_2+\mu_\mathcal{P}^{(3)}\mathfrak{l}^2+\mu_\mathcal{P}^{(3)}\mathfrak{u}_1^2+\frac{1}{2}\mathfrak{u}_1\dot{\mathfrak{l}}-\frac{1}{2}\mathfrak{l}\dot{\mathfrak{u}}_1+\dot{\mathfrak{u}}_2\\
        \chemV &= -2\mu_\mathcal{P}\mathfrak{v}_2-2\mu_\mathcal{J}\mathfrak{u}_2+2\mathfrak{l}\mathfrak{m}\mu_\mathcal{P}^{(3)}+2\mathfrak{u}_1\mathfrak{v}_1\mu_\mathcal{P}^{(3)}+\mu_\mathcal{J}^{(3)}\mathfrak{l}^2-\mu_\mathcal{J}^{(3)}\mathfrak{u}_1^2\nonumber\\
        &\quad+\frac{1}{2}\mathfrak{v}_1\dot{\mathfrak{l}}+\frac{1}{2}\mathfrak{u}_1\dot{\mathfrak{m}}-\frac{1}{2}\mathfrak{m}\dot{\mathfrak{u}}_1-\frac{1}{2}\mathfrak{l}\dot{\mathfrak{v}}_1+\dot{\mathfrak{v}}_2\,.
    \end{align}
    \end{subequations}
Consistency with the on-shell relations \eqref{eq:EOM} also requires that
    \begin{subequations}\label{eq:ChemPotRel}
    \begin{align}
        \mu_\mathcal{P} &= \chemL\mathcal{P}+\frac{8}{3}\chemU\mathcal{J}\mathcal{J}^{(3)}+\frac{4}{3}\chemU\mathcal{J}'-\frac{2}{3}\chemU'\mathcal{J}-\chemL'\\
        \mu_\mathcal{J} &= \chemM\mathcal{P}+\frac{8}{3}\chemU\mathcal{P}\mathcal{J}^{(3)}+\frac{8}{3}\chemU\mathcal{J}\mathcal{P}^{(3)}+\frac{8}{3}\chemV\mathcal{J}\mathcal{J}^{(3)}\nonumber\\
        &\quad+\frac{4}{3}\chemU\mathcal{P}'+\frac{4}{3}\chemV\mathcal{J}'-\frac{2}{3}\chemU'\mathcal{P}-\frac{2}{3}\chemV'\mathcal{J}-\chemM'\displaybreak[1]\\
        \mu_\mathcal{P}^{(3)} &= \chemL\mathcal{J}^{(3)}+\chemU\mathcal{J}^2-\frac{4}{3}\chemU\left(\mathcal{J}^{(3)}\right)^2-\chemU\mathcal{J}'-\frac{3}{2}\chemU'\mathcal{J}+\frac{1}{2}\chemU''\\
        \mu_\mathcal{J}^{(3)} &= \chemL\mathcal{P}^{(3)}+\chemM\mathcal{J}^{(3)}+2\chemU\mathcal{P}\mathcal{J}+\chemV\mathcal{J}^2-\frac{8}{3}\chemU\mathcal{P}^{(3)}\mathcal{J}^{(3)}-\frac{4}{3}\chemV\left(\mathcal{J}^{(3)}\right)^2\nonumber\\
        &\quad-\chemV\mathcal{J}'-\chemU\mathcal{P}'-\frac{3}{2}\chemV'\mathcal{J}-\frac{3}{2}\chemU'\mathcal{P}+\frac{1}{2}\chemV''\,.
    \end{align}
    \end{subequations}
The gauge fields $a$ and $\tilde{a}$ are then mapped to each other provided the following (twisted) Sugawara-like relations hold between the near horizon state-dependent functions $\mathcal{J}$, $\mathcal{P}$,  $\mathcal{J}^{(3)}$, $\mathcal{P}^{(3)}$ and the asymptotic state-dependent functions $\mathcal{M}$, $\mathcal{N}$, $\mathcal{V}$, $\mathcal{Z}$:
    \begin{subequations}\label{eq:FSMiuraTrafos}
    \begin{align}
        \mathcal{M} &= \mathcal{J}^2+\frac{4}{3}\left(\mathcal{J}^{(3)}\right)^2+2\mathcal{J}'\\
        \mathcal{N} &= \mathcal{J}\mathcal{P}+\frac{4}{3}\mathcal{J}^{(3)}\mathcal{P}^{(3)}+\mathcal{P}'\displaybreak[1]\\
        \mathcal{V} &= \frac{1}{54}\left(18\mathcal{J}^2\mathcal{J}^{(3)}-8\left(\mathcal{J}^{(3)}\right)^3+9\mathcal{J}'\mathcal{J}^{(3)}+27\mathcal{J}\mathcal{J}^{(3)'}+9\mathcal{J}^{(3)''}\right)\\
        \mathcal{Z} &= \frac{1}{36}\left(6\mathcal{J}^2\mathcal{P}^{(3)}-8\mathcal{P}^{(3)}\left(\mathcal{J}^{(3)}\right)^2+3\mathcal{P}^{(3)}\mathcal{J}'+3\mathcal{J}^{(3)}\mathcal{P}'\right.\nonumber\\
        &\qquad\quad\left.+9\mathcal{J}\mathcal{P}^{(3)'}+9\mathcal{P}\mathcal{J}^{(3)'}+12\mathcal{P}\mathcal{J}\mathcal{J}^{(3)}+3\mathcal{P}^{(3)''}\right)\,.
    \end{align}
    \end{subequations}
In addition one can explicitly check that the equations of motion
	\begin{subequations}\label{eq:StateTimeEvolution}
	\begin{align}
		 \dot{\cal M} &= - 2 \chemL''' + 2 {\cal M} \chemL^\prime +  {\cal M}^\prime \chemL  + 24 {\cal V} \chemU^\prime + 16 {\cal V}^\prime \chemU\\
 		\dot{\cal N} &=  \tfrac12\,\dot{\cal M}\big|_{L\to M} + 2 {\cal N} \chemL^\prime + {\cal N}^\prime \chemL + 24 {\cal Z} \chemU^\prime  + 16 {\cal Z}^\prime \chemU\displaybreak[1]\\
 		\dot{\cal V} &= \tfrac{1}{12}\,\chemU''''' - \tfrac{5}{12}\,{\cal M} \chemU''' - \tfrac58\, {\cal M}^\prime \chemU'' - \tfrac38\,{\cal M}'' \chemU^\prime  + \tfrac13\,{\cal M}^2 \chemU^\prime \nonumber \\
&\quad  - \tfrac{1}{12}{\cal M}''' \chemU  + \tfrac13\,{\cal M} {\cal M}^\prime \chemU + 3 {\cal V} \chemL^\prime  + {\cal V}^\prime \chemL\\
 		\dot{\cal Z} &=   \tfrac12\,\dot{\cal V}\big|_{L\to M} - \tfrac{5}{12}\, {\cal N} \chemU''' - \tfrac58\,{\cal N}^\prime \chemU''  - \tfrac38\,{\cal N}'' \chemU^\prime + \tfrac23\, {\cal M} {\cal N} \chemU^\prime  \nonumber \\
&\quad  - \tfrac{1}{12}\,{\cal N}''' \chemU + \tfrac13\,({\cal M} {\cal N})^\prime \chemU  + 3 {\cal Z} \chemL^\prime  + {\cal Z}^\prime \chemL\,,
	\end{align}
with
	\begin{align}
  		\tfrac12\,\dot{\cal M}\big|_{L\to M} &= -\chemM''' + {\cal M} \chemM^\prime  + \tfrac12\,{\cal M}^\prime(1+\chemM) + 12 {\cal V} \chemV^\prime + 8 {\cal V}^\prime \chemV \\
 		\tfrac12\,\dot{\cal V}\big|_{L\to M} &=  \tfrac{1}{24}\,\chemV''''' - \tfrac{5}{24}\,{\cal M} \chemV''' - \tfrac{5}{16}\,{\cal M}^\prime \chemV'' - \tfrac{3}{16}\,{\cal M}'' \chemV^\prime  + \tfrac16\,{\cal M}^2 \chemV^\prime  \nonumber \\
&\quad - \tfrac{1}{24}\,{\cal M}''' \chemV + \tfrac16\,{\cal M} {\cal M}^\prime \chemV + \tfrac32\,{\cal V} \chemM^\prime + \tfrac12\,{\cal V}^\prime(1+\chemM)\,,
	\end{align}
	\end{subequations}
indeed reduce to the simple ones given by \eqref{eq:EOM}. The relations \eqref{eq:ChemPotRel} show that the ``asymptotic chemical potentials'' $\chemL$, $\chemM$, $\chemU$, $\chemV$ depend not only on the ``near horizon chemical potentials'' $\mu_{\mathcal{P}}$, $\mu_{\mathcal{J}}$, $\mu_{\mathcal{P}}^{(3)}$, $\mu_{\mathcal{J}}^{(3)}$ but also on the state-dependent functions $\mathcal P$, $\mathcal J$, $\mathcal{P}^{(3)}$, $\mathcal{J}^{(3)}$, which is one way to see that our near horizon boundary conditions \eqref{eq:gaugeans}--\eqref{eq:FSCSpin3BCs} are inequivalent to the asymptotic ones of \cite{Afshar:2013vka, Gonzalez:2013oaa}. Moreover, the same relations directly map the corresponding gauge parameters that preserve the respective boundary conditions by replacing $\chemL\rightarrow\epsilon$, $(1+\chemM)\rightarrow\tau$, $\chemU\rightarrow\rchi$, $\chemV\rightarrow\kappa$ as well as $\mu_\mathcal{J}\rightarrow\epsilon_\mathcal{J}$, $\mu_\mathcal{P}\rightarrow\epsilon_\mathcal{P}$, $\mu_\mathcal{J}^{(3)}\rightarrow\epsilon_\mathcal{J}^{(3)}$ and $\mu_\mathcal{P}^{(3)}\rightarrow\epsilon_\mathcal{P}^{(3)}$. Therefore, also the infinitesimal transformation laws for $\mathcal{N}$, $\mathcal{M}$, $\mathcal{V}$ and $\mathcal{Z}$ can be directly read off from \eqref{eq:StateTimeEvolution} by replacing e.g. $\dot{\mathcal{M}}$ by  $\delta\mathcal{M}$ as well as all occurrences of chemical potentials $\mu_a$ and $\mu_a^{(3)}$ by the corresponding gauge parameters $\epsilon_a$ and $\epsilon^{(3)}_a$, respectively.

Thus, one can readily see that the fields $\mathcal{N}$, $\mathcal{M}$, $\mathcal{V}$ and $\mathcal{Z}$ transform exactly in such a way that they satisfy the $\mathcal{FW}_3$ algebra. Note, however, that their associated canonical charges still satisfy the (semidirect sum of four) $\hat{\mathfrak{u}}(1)$ current algebras as before. This can be seen by looking at the variation of the canonical boundary charge. In particular, after using the infinitesimal gauge transformations encoded in \eqref{eq:StateTimeEvolution}, the relations between the chemical potentials \eqref{eq:ChemPotRel} and the Miura-like transformations \eqref{eq:FSMiuraTrafos} reduces to
    \begin{align}
		\delta\mathcal{Q} &=\frac{k}{2\pi}\int\extd\varphi\left(\epsilon\,\delta\mathcal{N}+\frac{\tau}{2}\,\delta\mathcal{M}+8\rchi\,\delta\mathcal{Z}+4\kappa\,\delta\mathcal{V}\right)\nonumber\\
		&=\frac{k}{2\pi}\int\extd\varphi
		\left(
		\epsilon_\mathcal{J} \delta\mathcal{J} 
		+ \epsilon_\mathcal{P} \delta\mathcal{P}
		+ \frac{4}{3}\epsilon_\mathcal{J}^{(3)} \delta\mathcal{J}^{(3)}
		+\frac{4}{3} \epsilon_\mathcal{P}^{(3)} \delta\mathcal{P}^{(3)}
		\right)\,.
	\end{align}

\section{$\mathcal{FW}$-Algebras from Heisenberg}

In this section we relate the $\mathcal{FW}$-algebra to the near-horizon Heisenberg (or $\hat{\mathfrak{u}}(1)$ current) algebras. Using the (twisted) Sugawara-like relations \eqref{eq:FSMiuraTrafos} between the state-dependent functions as well as their Fourier mode expansions \eqref{eq:CurrentFourierModes} and
    \begin{subequations}
	\begin{align}
		\mathcal{N}(\varphi)&=\frac{1}{k}\sum\limits_{n\in\mathbb{Z}}L_ne^{-in\varphi}&
		\mathcal{M}(\varphi)&=\frac{2}{k}\sum\limits_{n\in\mathbb{Z}}M_ne^{-in\varphi}\\
		\mathcal{Z}(\varphi)&=\frac{\sqrt{3}}{8k}\sum\limits_{n\in\mathbb{Z}}U_ne^{-in\varphi}&
		\mathcal{V}(\varphi)&=\frac{\sqrt{3}}{4k}\sum\limits_{n\in\mathbb{Z}}V_ne^{-in\varphi}
	\end{align}
	\end{subequations}	
one finds that the (twisted) Sugawara construction for the $\mathcal{FW}_3$ algebra is given by
    \begin{subequations}\label{eq:FSMiuraTrafosFourier}
    \begin{align}
        L_n &= \frac{1}{k}\sum\limits_{p\in\mathbb{Z}}\left(J_{n-p}P_p+\frac{3}{4}J^{(3)}_{n-p}P^{(3)}_p\right)-inP_n\\
        M_n &= \frac{1}{2k}\sum\limits_{p\in\mathbb{Z}}\left(J_{n-p}J_p+\frac{3}{8}J^{(3)}_{n-p}J^{(3)}_p\right)-inJ_n\displaybreak[1]\\
        U_n &= \frac{\sqrt{3}}{k^2}\sum\limits_{p,q\in\mathbb{Z}}\left[\left(J_{n-p-q}J_p-\frac{3}{4}J^{(3)}_{n-p-q}J^{(3)}_p\right)J^{(3)}_q+2J_{n-p-q}J^{(3)}_pP_q\right]\nonumber\\
        &\quad-\frac{\sqrt{3}i}{2k}\sum\limits_{p\in\mathbb{Z}}\left[\left(3n-2p\right)J_{n-p}^{(3)}P_p+(n+2p)J_{n-p}P_p^{(3)}\right]-\frac{\sqrt{3}}{2}n^2P^{(3)}_n\\
        V_n &= \frac{\sqrt{3}}{k^2}\sum\limits_{p,q\in\mathbb{Z}}\left(J_{n-p-q}J_p-\frac{1}{4}J^{(3)}_{n-p-q}J^{(3)}_p\right)J^{(3)}_q-\frac{\sqrt{3}i}{2k}\sum\limits_{p\in\mathbb{Z}}(3n-2p)J^{(3)}_{n-p}J_p-\frac{\sqrt{3}}{2}n^2J^{(3)}_n\,.
    \end{align}
    \end{subequations}
At this point it is important to note that we already implicitly assumed some kind of normal ordering prescription for the constituents of the non-linear operators appearing in \eqref{eq:FSMiuraTrafosFourier}. The ordering prescription we chose is in accordance with the ones for induced representations as shown in \cite{Campoleoni:2016vsh}. Computing the commutation relations of these new operators we find that they satisfy the $\mathcal{FW}_3$ algebra \cite{Afshar:2013vka}
	\begin{subequations}\label{eq:FlatSpaceSpin3ASAFromU1}
		\begin{align}
		[L_n,L_m]=&(n-m)L_{n+m}\\
		[L_n,M_m]=&(n-m)M_{n+m}+\frac{c_M}{12}n(n^2-1)\delta_{n+m,0}\displaybreak[1]\\
		[L_n,U_m]=&(2n-m)U_{n+m}\\
		[L_n,V_m]=&(2n-m)V_{n+m}\\
		[M_n,U_m]=&(2n-m)V_{n+m}\displaybreak[1]\\
		[U_n,U_m]=&-\frac{1}{3}(n-m)(2n^2+2m^2-nm-8)L_{n+m}-\frac{64}{c_M}(n-m)\Lambda_{n+m}\\
		[U_n,V_m]=&-\frac{1}{3}(n-m)(2n^2+2m^2-nm-8)M_{n+m}-\frac{32}{c_M}(n-m)\Theta_{n+m}\nonumber\\
				&-\frac{c_M}{36}n(n^2-4)(n^2-1)\delta_{n+m,0}\,,
		\end{align}
	\end{subequations}
with
	\begin{equation}
		\Lambda_{n}=\sum_{p\in\mathbb{Z}}M_pL_{n-p},\qquad\Theta_n=\sum_{p\in\mathbb{Z}}M_pM_{n-p}
	\end{equation}
and $c_M=12k$. In addition the spin-2 and spin-3 generators have the following non-vanishing commutation relations with the spin-1 currents:
	\begin{subequations}\label{eq:FlatSpaceSpin3ASAAndU1}
		\begin{align}
		[L_n,P_m]=&-mP_{n+m}\\
		[L_n,J_m]=&-mJ_{n+m}-in^2k\delta_{n+m,0}\\
		[M_n,P_m]=&-mJ_{n+m}-in^2k\delta_{n+m,0}\displaybreak[1]\\
		[L_n,P^{(3)}_m]=&-mP^{(3)}_{n+m}\\
		[L_n,J^{(3)}_m]=&-mJ^{(3)}_{n+m}\\
		[M_n,P^{(3)}_m]=&-mJ^{(3)}_{n+m}\displaybreak[1]\\
		[U_n,P_m]=&-\frac{2\sqrt{3}}{k}m\sum\limits_{p\in\mathbb{Z}}\left(J_{n+m-p}P^{(3)}_q+J^{(3)}_{n+m-p}P_q\right)+\frac{\sqrt{3}i}{2}m(3n+2m)P^{(3)}_{n+m}\\
		[U_n,J_m]=&-\frac{2\sqrt{3}}{k}m\sum\limits_{p\in\mathbb{Z}}J_{n+m-p}J^{(3)}_q+\frac{\sqrt{3}i}{2}m(3n+2m)J^{(3)}_{n+m}\\
		[V_n,P_m]=&-\frac{2\sqrt{3}}{k}m\sum\limits_{p\in\mathbb{Z}}J_{n+m-p}J^{(3)}_q+\frac{\sqrt{3}i}{2}m(3n+2m)J^{(3)}_{n+m}\displaybreak[1]\\
		[U_n,P^{(3)}_m]=&\frac{2\sqrt{3}}{k}m\sum\limits_{p\in\mathbb{Z}}\left(J^{(3)}_{n+m-p}P^{(3)}_q-\frac{4}{3}J_{n+m-p}P_q\right)+\frac{2i}{\sqrt{3}}m(3n+2m)P_{n+m}\\
		[U_n,J^{(3)}_m]=&\frac{\sqrt{3}}{k}m\sum\limits_{p\in\mathbb{Z}}J^{(3)}_{n+m-p}J^{(3)}_q+\frac{2i}{\sqrt{3}}m(3n+2m)J_{n+m}-\frac{2k}{\sqrt{3}}n^3\delta_{n+m,0}\\
		[V_n,P^{(3)}_m]=&\frac{\sqrt{3}}{k}m\sum\limits_{p\in\mathbb{Z}}J^{(3)}_{n+m-p}J^{(3)}_q+\frac{2i}{\sqrt{3}}m(3n+2m)J_{n+m}-\frac{2k}{\sqrt{3}}n^3\delta_{n+m,0}\,.
		\end{align}
	\end{subequations}


\section{Entropy of cosmological solutions}
\label{blackholeentropy}

A flat space cosmology (FSC) is described by the field configuration \eqref{eq:FSCSpin3Connection} with $\mathcal{V}=\mathcal{Z}=\mu_L=\mu_M=\mu_U=\mu_V=0$ and constant $\mathcal{M}$, $\mathcal{N}$. 
The entropy of a FSC with inverse temperature $\beta=\frac{1}{T}$, angular velocity $\Omega$, energy $H$ and angular momentum $J$ satisfies a first law \cite{Bagchi:2012xr} that is given by
    \begin{equation}\label{eq:FirstLaw}
        \delta H=-T\delta S+\Omega\delta J\,.
    \end{equation}
In \eqref{eq:HamiltonianVar} we already computed $\delta H$. For $\delta J$ one can proceed in exactly the same way i.e.
    \begin{equation}
        \delta J:=\delta Q[\partial_\varphi]=\frac{k}{2\pi}\int\extd\varphi\langle\xi^\varphi\mathcal{A}_\varphi\delta\mathcal{A}_\varphi\rangle=\frac{k}{2\pi}\int\extd\varphi\langle\mathcal{A}_\varphi\delta\mathcal{A}_\varphi\rangle\,.
    \end{equation}
Thus one can rewrite \eqref{eq:FirstLaw} also as
    \begin{equation}\label{eq:VariationEntropy}
        \delta S=-\frac{k}{2\pi}\beta\int\extd\varphi \langle a_v\delta a_\varphi\rangle+\frac{k}{2\pi}\beta\,\Omega\int\extd\varphi \langle a_\varphi\delta a_\varphi\rangle\,.
    \end{equation}
As in the AdS$_3$ case we impose that the holonomy of $h=-i\frac{\beta}{2\pi}\left(\int\extd\varphi\,a_v - \Omega \int\extd\varphi\,a_\varphi\right)$ is in the center of the gauge group i.e. 
    \begin{equation}
        \textrm{Eigen}\left[h\right]\propto \textrm{Eigen}\left[L_0\right]\,.
    \end{equation}
In order to make contact with the thermal entropy of FSCs we demand that the holonomies of our boundary conditions match the ones of FSCs. That means the holonomies are fixed to
     \begin{equation}\label{eq:HolCond}
        \textrm{Eigen}\left[h\right]= \textrm{Eigen}\left[2\pi i L_0\right]\,.
    \end{equation}   
Assuming again constant chemical potentials for the boundary conditions \eqref{eq:FSCSpin3BCs} the holonomy conditions \eqref{eq:HolCond} yield the following restrictions
    \begin{equation}
        \mu_\mathcal{P}=-\frac{4\pi^2}{\beta}+J_0\Omega,\quad\mu_\mathcal{J}=P_0\Omega,\quad\mu_\mathcal{P}^{(3)}=\Omega J_0^{(3)},\quad\mu_\mathcal{J}^{(3)}=\Omega P_0^{(3)}\,.
    \end{equation}
Using these conditions the variation of the entropy \eqref{eq:VariationEntropy} simplifies considerably and can be functionally integrated to yield
    \begin{equation}\label{eq:HairyEntropy}
        S_{\textrm{Th}}=2\pi P_0= 2\pi (J_0^{+}+J_{0}^{-})\,.
    \end{equation}
In order to relate this entropy formula with the one for the higher-spin case one first has to solve \eqref{eq:FSMiuraTrafos} for $P_0$. Introducing the dimensionless ratios\footnote{The real numbers $N$, $M$, $V$ and $Z$ denote the zero modes of the functions $\mathcal{N}$, $\mathcal{M}$, $\mathcal{V}$ and $\mathcal{Z}$ respectively.}
	\begin{equation}
		\frac{V}{2M^{\frac{3}{2}}}=\frac{\mathcal{R}-1}{\mathcal{R}^\frac{3}{2}}\qquad \textrm{and} \qquad \frac{Z}{N\sqrt{M}}=\mathcal{P}\,,
	\end{equation}
the solution for $P_0$ in terms of these ratios reads
    \begin{equation}
        P_0=k\frac{N\left(4\mathcal{R}-6+3\mathcal{P}\sqrt{\mathcal{R}}\right)}{4\sqrt{M}(\mathcal{R}-3)\sqrt{1-\frac{3}{4\mathcal{R}}}}\,.
    \end{equation}
Plugging this expression into \eqref{eq:HairyEntropy} one immediately obtains
    \begin{equation}
      \label{eq:Sbound}
        S_{\textrm{Th}}=2\pi k\frac{ N\left(2\mathcal{R}-6+3\mathcal{P}\sqrt{\mathcal{R}}\right)}{8\sqrt{M}(\mathcal{R}-3)\sqrt{1-\frac{3}{4\mathcal{R}}}}\,,
    \end{equation}
which is exactly the entropy of a FSC with spin-3 hair and central charge $c_M=12k$ \cite{Gary:2014ppa,Matulich:2014hea,Riegler:2014bia,Basu:2015evh,Riegler:2016hah}.

\section{Metric formulation}
\label{se:metric}
In this section we present some of our results in the metric formulation for convenience of readers more familiar with that formulation. While this means that we merely translate results from previous sections, it can be useful for future applications to have explicit expressions for metric and spin-3 field, see \cite{Campoleoni:2012hp, Campoleoni:2014tfa} for some reasons to consider the metric formulation and for AdS$_3$ results. For instance, if one wants to add matter couplings the Chern--Simons formulation loses some of its attractiveness, while the metric formulation remains suitable \cite{Prokushkin:1998bq}. Even though no non-linear action is known in this formulation, even perturbatively some non-trivial cross-checks are possible, like a comparison of Wald's entropy with our result \eqref{eq:HairyEntropy} to quadratic order in the spin-3 field, along the lines of \cite{Campoleoni:2012hp}. Finally, for some flat-space holographic purposes it can be useful to have the metric formulation available, e.g.~for the identification of sources and vacuum expectation values, see \cite{Gary:2014ppa} and references therein.

We start now with the translation of our results into the metric formulation, assuming for simplicity constant chemical potentials.
The metric is the contraction over the local zuvielbein, which can be extracted from the connection \eqref{eq:gaugeans}
via 
\begin{equation}
\label{metricfromA}
\extd s^2 = - 2 \,\eta_{m n} \mathcal{A}_{\Mt}^m \mathcal{A}_{\Mt}^n + \frac{2}{3} \,\mathcal{K}_{m n} \mathcal{A}_{\Vt}^m \mathcal{A}_{\Vt}^n
\end{equation}
with $\eta_{m n} = \mathrm{diag}(1, -1/2, 1)$ and $\mathcal{K}_{m n} = \mathrm{diag}(12, -3, 2, -3, 12)$ and $\mathcal{A}_{\Mt}^m,  \mathcal{A}_{\Vt}^n$ denoting the coefficients of the connection with respect to the spin-2 and spin-3 translation operators $\Mt_m$ and $\Vt_n$.  Using the twisted and hatted traces in combination with the matrix representations as defined in \cite{Gary:2014ppa} one can rewrite \eqref{metricfromA} as
\begin{equation}
\extd s^2 = \tfrac{1}{2}\, \tilde{\mathrm{tr}}\left( A_{\mu} A_{\nu} \right) \extd x^\mu\, \extd x^\nu\,.
\end{equation}
In the same gauge, the spin-3 field can be computed using 
again the twisted trace as
\begin{equation}
 \label{spin3fielddef}
 \Phi = \tfrac{1}{6}\,  \tilde{\mathrm{tr}}\left( A_{\mu} A_{\nu} A_{\lambda} \right) \extd x^\mu\, \extd x^\nu \extd x^{\lambda}\,.
\end{equation}

Using the definitions above together with \eqref{eq:gaugeb} and \eqref{eq:FSCSpin3BCs} yields the metric
\begin{align}
\label{metric}
\dd s^2 &=  2 \left(\text{dv}+\frac{ \mathcal{J}}{\mu_\mathcal{\mathcal{P}}}\text{d$\varphi $}\right) \text{dr}
            + \left(\mu _{\mathcal{J}}^2+\frac{4 }{3} (\mu^{(3)}_\mathcal{P})^2 +2 r
   \mu _\mathcal{P} +\frac{8 r (\mu^{(3)} _{\mathcal{J}})^2}{\mu _\mathcal{\mathcal{P}}} \right) \text{dv}^2\nonumber\\
   &\quad +2 \left(\mathcal{P} \mu _\mathcal{J} 
     +\frac{4}{3} \mathcal{P}^{(3)} \mu^{(3)}_\mathcal{P}
     +2 \mathcal{J} r 
     + \frac{8 \mathcal{J}^{(3)} r \mu^{(3)}_{\mathcal{J}}}{\mu_\mathcal{P}}\right) \text{dv}\, \text{d$\varphi $} \nonumber \\
   &\quad+ \left(\mathcal{P}^2
     +\frac{4}{3} (\mathcal{P}^{(3)})^2
     + \frac{2 \mathcal{J}^2 r}{\mu_\mathcal{P}}
     +\frac{8 (\mathcal{J}^{(3)})^2 r}{\mu_\mathcal{P}}\right) \text{d$\varphi $}^2 \,
\end{align}
and the spin-3 field
\begin{align}
\label{spin3}
\Phi = \frac{2}{27 \mu_\mathcal{P}} \,&\big(\Phi_{vvv}\,\extd v^3 + \Phi_{vv\varphi}\,\extd v^2\extd\varphi + \Phi_{vvr}\,\extd v^2 \extd r + \Phi_{v\varphi r} \,\extd v \extd\varphi\extd r \nonumber \\ 
 &  + \Phi_{v\varphi\varphi}\,\extd v \extd\varphi^2+ \Phi_{\varphi\varphi\varphi}\,\extd\varphi^3
+ \Phi_{\varphi\varphi r}\,\extd\varphi^2\extd r \big)
\end{align}
the components of which are given in appendix \ref{app:C}.

\section{Extension to fields with spin greater than three}
\label{sec:extens-high-spin}

In this section we discuss the generalization to fields of spin $s=2,3, \ldots, N$, i.e., the algebra of our Chern-Simons theory is $\mathfrak{isl}(N,\R)$ (see appendix \ref{sec:ihsisl} for our conventions).
The following calculation works equally well for $\mathfrak{ihs}[\lambda]$.
We use the same gauge as in the aforementioned cases, see in particular equations \eqref{eq:gaugeans} and \eqref{eq:gaugeb}, and propose the boundary conditions
    \begin{subequations}\label{eq:FSCSpinNBCs}
    \begin{align}
      a_\varphi&=a_\varphi^{(2,3)}+ \sum_{s=4}^{N} \mathcal{J}^{(s)} \, \Lt_0^{(s)}
                 + \sum_{s=4}^{N} \mathcal{P}^{(s)} \, \Mt_0^{(s)}
                 =\sum_{s=2}^{N}\left( \mathcal{J}^{(s)} \, \Lt_0^{(s)} + \mathcal{P}^{(s)} \, \Mt_0^{(s)} \right) \, , \\
      a_v&=a_v^{(2,3)} + \sum_{s=4}^{N} \chemP^{(s)} \, \Lt_0^{(s)}
        + \sum_{s=4}^{N} \chemJ^{(s)}  \, \Mt_{0}^{(s)} 
           =\sum_{s=2}^{N} \left(\chemP^{(s)} \, \Lt_0^{(s)} +\chemJ^{(s)}  \, \Mt_{0}^{(s)} \right) \, ,
    \end{align}
    \end{subequations}
where $a^{(2,3)}$ refers to the spin-2 and spin-3 part of the connection used in the previous sections. The considerations of section \ref{nearhorizon} generalize.  
This is connected to the fact that the generators $\Lt_{0}^{(s)}$ and $\Mt_{0}^{(t)}$ commute among themselves and with each other also for the higher $N$ cases.  
The canonical boundary charge is then 
\begin{align}
  Q[\epsilon]= \frac{k}{2\pi}\int\extd\varphi 
  \sum_{s=2}^{N} \alpha_{s} 
  \left(
  \epsilon_\mathcal{J}^{(s)} \mathcal{J}^{(s)}+ \epsilon_\mathcal{P}^{(s)} \mathcal{P}^{(s)}
  \right)
\end{align}
and the corresponding asymptotic symmetry algebra is given by 
    \begin{equation}\label{eq:FourierASAN}
      [J_n^{(s)},P^{(t)}_m]=\alpha_{s} k \, n \, \delta^{s,t}\delta_{n+m,0} \,.
    \end{equation} 
The constants $\alpha_{s}$ are coming from the invariant metric of the $\mathfrak{isl}(N,\R)$ algebra, see equation \eqref{eq:invmeihs}.
Relating the near horizon boundary conditions above to the asymptotic ones is a conceptually straightforward technical problem that we will not address here. 

An interesting aspect of our near horizon boundary conditions is that as in the AdS$_{3}$ higher-spin case the entropy calculated in terms of the near horizon boundary conditions is unchanged under the addition of the higher-spin fields and still given by
\begin{align}
  S_{\mathrm{Th}}= 2 \pi P_{0} = 2 \pi (J_{0}^{+} + J_{0}^{-}) \,.
\end{align}
As for the spin-3 case we focused here on the branch that is continuously connected to the FSC.

\section{Conclusions}
\label{conclusions}

We have proposed new boundary conditions for flat space spin-3 gravity \eqref{eq:gaugeans}-\eqref{eq:FSCSpin3BCs} and flat space higher-spin gravity \eqref{eq:FSCSpinNBCs}, and have shown that they lead to well defined charges as well as to a novel asymptotic symmetry algebra, \eqref{eq:FourierASA} and \eqref{eq:FourierASAN}, respectively. Using this algebra we have shown in section \ref{sec:SoftHair} how soft excitations can be created by acting with its generators on some states, like the vacuum or flat space cosmologies. The relation between the near horizon and the asymptotic quantities given in equation \eqref{eq:FSMiuraTrafos} made it possible to relate the remarkably simple entropy of the near horizon geometries, \eqref{eq:HairyEntropy}, to the more complicated one from the boundary, \eqref{eq:Sbound}.

This work shows another generalization of the proposal of \cite{Afshar:2016wfy}. Interestingly, the generalization to supersymmetric versions~\cite{Achucarro:1986vz}, non-principally embedded (higher) spins~\cite{Ammon:2011nk,Campoleoni:2011hg,Castro:2012bc} as well es higher dimensions has not yet been achieved and might therefore provide a further testing ground. Some of these generalizations are technically (and perhaps also conceptually) more challenging, since even for situations where a Chern--Simons formulation is available the connection most likely is not going to be diagonal, as opposed to the situation in previous work or in our current paper.

Our considerations have focused on the (higher-spin generalization) of future null infinity. Investigations using the full structure of asymptotically flat spacetimes~\cite{Strominger:2013jfa} have provided fascinating connections between soft modes and conservation laws~\cite{He:2014laa,Cachazo:2014fwa,Strominger:2014pwa}.
In the light that future and null infinity have also been linked in three spacetime dimensions~\cite{Prohazka:2017equ} it might be fruitful to search for similar effects for the flat space case of \cite{Afshar:2016kjj} as well as for our proposal.
Progress in this direction in four spacetime dimensions including higher-spin extensions has been made in \cite{Campoleoni:2017mbt}.

Finally, an explicit construction of flat space cosmology microstates, along the lines of the corresponding BTZ construction \cite{Afshar:2016uax, Sheikh-Jabbari:2016npa} is an outstanding open problem (both for spin-2 and higher-spins) that might even provide insights into the black hole information paradox~\cite{Hawking:2016msc}.

\section*{Acknowledgments}
We thank Wout Merbis for insightful discussions and Hamid Afshar, St{\'e}phane Detournay, Alfredo Perez, David Tempo, Ricardo Troncoso, Shahin Sheikh-Jabbari and Hossein Yavartanoo for collaboration on other aspects of soft Heisenberg hair.

DG was supported by the FWF projects P 27182-N27 and P 28751-N2 and during the final stage also by the program Science without Borders, project CNPq-401180/2014-0.
SP was supported by the FWF project P 27396-N27. The research of MR is supported by the ERC Starting Grant 335146 ``HoloBHC''.

\appendix

\section{$\mathfrak{isl}(3,\mathbb{R})$  algebra}
\label{append}
Here we review our conventions for the algebra $\mathfrak{isl}(3,\mathbb{R})$. The set of generators that span the $\mathfrak{isl}(3,\mathbb{R})$ algebra 
is given by ${\Lt}_i, {\Mt}_i, {\Ut}_m, {\Vt}_m$ 
with $i = -1, 0, 1$ and $m = -2, -1, 0, 1, 2$ such that
\begin{subequations}
\label{isl3Ralgebraexplicit}
\begin{align}
[\Lt_n, \Lt_m] & = (n - m) \Lt_{n+m}
\\
[\Lt_n, \Mt_m] &= (n-m) \Mt_{n+m}
\\
[\Lt_n, \Ut_m] &= (2n-m) \Ut_{n+m}
\\
[\Lt_n, \Vt_m] &= (2n-m) \Vt_{n+m}
\\
[\Ut_n, \Ut_m] &= \sigma (n-m) (2 n^2 + 2 m^2 -nm -8) \Lt_{n+m}
\\
[\Ut_n, \Vt_m] &= \sigma (n-m) (2 n^2 + 2 m^2 -nm -8) \Mt_{n+m}\,.
\end{align}
\end{subequations}
The $\Lt_n$ generate rotations, the $\Mt_n$ generate translations and $\Ut_n, \Vt_n$ generate spin-3 transformations. The factor $\sigma$ fixes the normalization of the spin-3 generators $\Ut_n$ and $\Vt_n$. We choose
\begin{equation}
\label{sigmavalue}
\sigma = - \frac{1}{3}\, .
\end{equation}
This algebra may be equipped with an invariant metric, which is a non-degenerate, invariant, symmetric bilinear form, given by 
	\begin{equation}\label{eq:ISLInvBilForm}
		\langle \Lt_n \, \Mt_m\rangle=-2\left(
			\begin{array}{c|ccc}
				  &\Mt_1&\Mt_0&\Mt_{-1}\\
				\hline
				\Lt_1 & 0 & 0 & 1\\
				\Lt_0 & 0 & -\frac{1}{2} & 0\\
				\Lt_{-1} & 1 & 0 & 0
			\end{array}\right) 
	\end{equation}
as well as
	\begin{equation}\label{eq:ISLInvBilFormSpin3}
		\langle \Ut_n \, \Vt_m\rangle=2\left(
			\begin{array}{c|ccccc}
				& \Vt_2 & \Vt_1 & \Vt_0 & \Vt_{-1} & \Vt_{-2}\\
				\hline
				\Ut_2 & 0 & 0 & 0 & 0 & 4\\
				\Ut_1 & 0 & 0 & 0 & -1 & 0\\
				\Ut_0 & 0 & 0 & \frac{2}{3} & 0 & 0\\
				\Ut_{-1} & 0 & -1 & 0 & 0 & 0\\
				\Ut_{-2} & 4 & 0 & 0 & 0 & 0
			\end{array}\right) \, .
	\end{equation}
All other pairings of generators inside the bilinear form are zero.

\section{$\mathfrak{ihs}[\lambda]$ and $\mathfrak{isl}(N,\R)$  algebra}
\label{sec:ihsisl}

We define the infinite dimensional Lie algebra $\mathfrak{ihs}[\lambda]$ using a contraction of $\mathfrak{hs}[\lambda] \oplus \mathfrak{hs}[\lambda]$.
The finite dimensional algebra $\mathfrak{isl}(N,\R)$ is then given by a Lie algebra quotient thereof.
We will provide an invariant metric for both algebras as well as the commutators for spins $s \leq 4$ of $\mathfrak{ihs}[\lambda]$.

\subsection{Contraction}
\label{sec:contraction}

Here we will sketch how $\mathfrak{ihs}[\lambda]$ can be derived as a contraction of $\mathfrak{hs}[\lambda] \oplus \mathfrak{hs}[\lambda]$\footnote{%
This construction works equally well for any other direct sum of a Lie algebra with an invariant metric.}.
A basis for the first and second summand are given by $\Et_{a}$ and $\tilde\Et_{a}$, respectively. The commutation relations are then (we suppress various indices for clarity)
\begin{align}
  [\Et_{a},\Et_{b}]&=f\indices{_{ab}^{c}} \Et_{c}
&
  [\Et_{a},\tilde \Et_{b}]&=0
&
  [\tilde\Et_{a},\tilde\Et_{b}]&=f\indices{_{ab}^{c}} \tilde \Et_{c} 
\end{align}
and the invariant metric is
\begin{align}
  \langle \Et_{a} \Et_{b} \rangle &= \Omega_{ab}
&
  \langle \Et_{a} \tilde \Et_{b} \rangle &= 0
&
  \langle \tilde\Et_{a} \tilde\Et_{b} \rangle = \Omega_{ab} \, .
\end{align}
Defining
\begin{align}
  \Lt_{a}&=\Et_{a}+\tilde \Et_{a}
&
  \Mt_{a}&=\frac{1}{\ell}(\Et_{a}-\tilde \Et_{a})
\end{align}
and taking the  $\ell \to \infty$ limit leads to the new algebra
\begin{align}
  [\Lt_{a},\Lt_{b}]&=f\indices{_{ab}^{c}} \Lt_{c}
&
  [\Lt_{a}, \Mt_{b}]&=f\indices{_{ab}^{c}} \Mt_{c}
&
  [\Mt_{a},\Mt_{b}]&=0
\end{align}
with the invariant metric $\langle \Lt_{a} \Mt_{b} \rangle = \Omega_{ab}$ and  $\langle \Lt_{a} \Lt_{b} \rangle = \langle \Mt_{a} \Mt_{b} \rangle =0$. The invariance of $\langle \Lt_{a} \Mt_{b} \rangle$ with respect to $\Lt_{c}$ is a consequence of the invariance of the original invariant metric.

\subsection{$\mathfrak{ihs}[\lambda]$}
\label{sec:ihs}

The generators of $\mathfrak{ihs}[\lambda]$ are given by 
\begin{align}
  \Lt_{n}^{(s)},\Mt_{n}^{(s)}, \quad  s\geq2 , \quad |n|<s \,.
\end{align}
With the notation used in the previous sections $\Lt_{n}^{(2)}=\Lt_{n}$, $\Mt_{n}^{(2)}=\Mt_{n}$, $\Lt_{n}^{(3)}=\Ut_{n}$ and $\Mt_{n}^{(3)}=\Vt_{n}$.
Using the contraction described in the preceding subsection we can use the commutation relations of $\mathfrak{hs}[\lambda]$~\cite{Feigin:1988,Bordemann:1989zi,Bergshoeff:1989ns,Pope:1989sr,Fradkin:1990ir}\footnote{%
The commutation relations were explicitly given in \cite{Pope:1989sr}. Our structure constants are divided by four with respect to the ones given in \cite{Gaberdiel:2011wb}, but we otherwise closely follow \cite{Gaberdiel:2011wb} (see also \cite{Henneaux:2010xg,Campoleoni:2011hg,Ammon:2011ua}).
}
we arrive at the commutation relations of $\mathfrak{ihs}[\lambda]$ \cite{Riegler:2016hah}
\begin{subequations}
\label{eq:ihs}
  \begin{align}
    [\Lt_{n}^{(s)},\Lt_{m}^{(t)}]&=\sum_{ \stackrel{u=2}{\mathrm{ even}}}^{s+t-1}  g_u^{st}(n,m;\lambda) \,\Lt_{n+m}^{(s+t-u)}
    \\
    [\Lt_{n}^{(s)},\Mt_{m}^{(t)}]&=\sum_{ \stackrel{u=2}{\mathrm{ even}}}^{s+t-1}  g_u^{st}(n,m;\lambda) \,\Mt_{n+m}^{(s+t-u)}
    \\
    [\Mt_{n}^{(s)},\Mt_{m}^{(t)}]&=0  
  \end{align}
\end{subequations}
where
\begin{subequations}
  \begin{align}
    g_u^{st}(n,m;\lambda) &= {\frac{q^{u-2}}{2(u-1)!}} \phi_u^{st}(\lambda) N_u^{st}(n,m)
                            \\
    N_u^{st}(n,m) &=  \sum_{k=0}^{u-1}(-1)^k
                    \binom{u-1}{k}
                    [s-1+n]_{u-1-k}[s-1-n]_k[t-1+m]_k[t-1-m]_{u-1-k}
    \\
    \phi_u^{st}(\lambda) &= \ _4F_3\left[
                           \begin{array}{c|}
                             \frac{1}{2} + \lambda \ ,  \ \frac{1}{2} - \lambda \ , {\frac{2-u}{2}} \ , {\frac{1-u}{2}}\\
                             {\frac{3}{2}}-s \ , \ {\frac{3}{2}} -t\ , \ \frac{1}{2} + s+t-u
                           \end{array}  \ 1\right] \, .  
  \end{align}
\end{subequations}
The number $q$ is a normalization factor that can be set to any fixed value (for more details see Appendix A in \cite{Gaberdiel:2011wb}).
The falling factorial or Pochhammer symbol is given by 
\begin{align}
[a]_{n}= a (a-1) (a-2) \cdots (a-n+1) =\frac{a!}{(a-n)!} = \frac{\Gamma(a+1)}{\Gamma(a+1-n)}
\end{align}
the rising factorial or Pochhammer symbol is given by 
\begin{align}
  (a)_{n}= a (a+1) \cdots (a+n-1)= \frac{(a+n-1)!}{(a-1)!}= \frac{\Gamma(a+n)}{\Gamma(a)}
\end{align}
with
\begin{align}
  (a)_{0}=[a]_{0}=1 \,.
\end{align}
The generalized hypergeometric function $_{m}F_{n}(z)$ is defined by 
\begin{align}
  _{m}F_{n}
  \left[
  \begin{array}{c|}
    a_{1},\dots,a_{m} \\
    b_{1},\dots,b_{n}
  \end{array} \  z
  \right]
  = \sum_{k=0}^{\infty}\frac{(a_{1})_{k} (a_{2})_{k} \dots (a_{m})_{k}}{(b_{1})_{k} (b_{2})_{k} \dots (b_{n})_{k}} \frac{z^{k}}{k!} \,.
\end{align}
The infinite dimensional Lie algebra $\mathfrak{ihs}[\lambda]$ possesses an invariant metric given by 
\begin{subequations}
\label{eq:invmeihs}
  \begin{align}
    \langle \Lt^{(s)}_n \Mt^{(t)}_m \rangle &\equiv \frac{3}{4 q (\lambda^2-1)} g^{s t}_{s+t-1}(n,m,\lambda)\\
                                            &=N_s  \frac{(-1)^{s-n-1}}{4 (2s-2)!}\Gamma(s+n)\Gamma(s-n)  \delta^{st}\delta_{n,-m} \notag
    \\
    \langle \Lt^{(s)}_n \Lt^{(t)}_m \rangle&=\langle \Mt^{(s)}_n \Mt^{(t)}_m \rangle=0
  \end{align}
\end{subequations}
with
\begin{align}
    N_s &\equiv {3 \cdot 4^{s-3}\sqrt{\pi}q^{2s-4}\Gamma(s)\over (\lambda^2-1)
          \Gamma(s+\frac{1}{2})} (1-\lambda)_{s-1} (1+\lambda)_{s-1} \ . \label{appN}
\end{align}
The overall constant has been chosen so that
\begin{align}
\langle \Lt^{(2)}_1 \Mt^{(2)}_{-1} \rangle= -1 \,.
\end{align}

\subsection{$\mathfrak{isl}(N,\R)$}
\label{sec:islN}

Using $\mathfrak{ihs}[\lambda]$ one can define $\mathfrak{isl}(N,\R)$ as a Lie algebra quotient. This is only possible for $\lambda=N$ since this leads to an ideal $\chi_{N}$~\cite{Feigin:1988,Vasiliev:1989re,Fradkin:1990qk} spanned by $\Lt^{(s)}_{n}$ and $\Mt^{(s)}_{n}$ with $s>N$. Using this ideal we can then define the finite dimensional algebra $\mathfrak{isl}(N,\R)$ by the quotient
\begin{align}
  \mathfrak{isl}(N,\R)=\mathfrak{ihs}[N]/\chi_{N} \,.
\end{align}
The invariant metric, equation \eqref{eq:invmeihs} with $\lambda=N$, stays an invariant metric for $\mathfrak{isl}(N,\R)$. It is  zero for higher spins. In the next section this can be seen explicitly.

\subsection{Commutators of $\mathfrak{ihs}[\lambda]$ for $s \leq 4$}
\label{sec:first-commutators}

We list here the commutators for $s \leq 4$ of $\mathfrak{ihs}[\lambda]$ (with $q=1/4$)\footnote{%
A \texttt{Mathematica} workbook that reproduces the commutation relations and might be useful for further checks is uploaded as an ancillary file on the \texttt{arxiv} server.
}
\begin{subequations}
  \begin{align}
    [\Lt_{n}^{(2)},\Lt_{m}^{(2)}]&=(n-m) \Lt^{(2)}_{n+m}
    \\
    [\Lt_{n}^{(2)},\Lt_{m}^{(3)}]&=(2n-m) \Lt^{(3)}_{n+m}
    \\
    [\Lt_{n}^{(3)},\Lt_{m}^{(3)}]&=-\frac{1}{60}(\lambda^{2}-4) (n-m) (2 n^{2}-n m + 2m^{2} - 8 )  \Lt^{(2)}_{n+m}
                                   \nonumber \\
                                 &\quad + 2 (n-m) \Lt^{(4)}_{n+m}
    \\
    [\Lt_{n}^{(2)},\Lt_{m}^{(4)}]&=(3n-m) \Lt^{(4)}_{n+m} 
    \\
    [\Lt_{n}^{(3)},\Lt_{m}^{(4)}]&=-\frac{1}{70} (\lambda^{2}-9) (5 n^{3}-5n^{2}m -17 n + 3n m^{2} + 9 m - m^{3}) \Lt^{(3)}_{n+m} 
                                   \nonumber \\
                                 & \quad + (3 n - 2m) \Lt^{(5)}_{n+m}
    \\
    [\Lt_{n}^{(4)},\Lt_{m}^{(4)}]&=(\lambda^{2}-4)(\lambda^{2}-9) (n-m) f(n,m) \Lt^{(2)}_{n+m}
                                   \nonumber  \\
                                 & \quad -\frac{1}{30}(\lambda^{2}-19) (n-m) (n^{2}-nm + m^{2}-7) \Lt^{(4)}_{n+m}
                                   \nonumber  \\
                                 & \quad +3 (n-m) \Lt^{(6)}_{n+m}
  \end{align}
\end{subequations}
with
\begin{equation}
 f(n,m)=\frac{1}{8400}\left[3 n^4 + 3m^4 -2 n m (n-m)^2  - 39n^2 -39 m^2 + 20 n m + 108\right] \,.
\end{equation}
The commutators for $[\Lt_{n}^{(s)},\Mt_{m}^{(t)}]$ are equivalent with the substitution $\Lt \to \Mt$ on the right hand side [see equation \eqref{eq:ihs}].
The invariant metric for $s \leq 4$ is given by the anti-diagonal matrices
\begin{subequations}
  \begin{align}
    \langle \Lt^{(2)}_{n} \, \Mt^{(2)}_{m} \rangle &= \mathrm{adiag}(-1,\tfrac{1}{2},-1)
    \\
    \langle \Lt^{(3)}_{n} \, \Mt^{(3)}_{m} \rangle &= \frac{1}{20} (\lambda^{2}-4)\cdot \mathrm{adiag}(4,-1,\tfrac{2}{3},-1,4)
    \\
    \langle \Lt^{(4)}_{n} \, \Mt^{(4)}_{m} \rangle &= \frac{1}{140} (\lambda^{2}-4)(\lambda^{2}-9) \cdot \mathrm{adiag}(-6,1,\tfrac{2}{5}, \tfrac{3}{10}, \tfrac{2}{5},1,-6) \,.
  \end{align}
\end{subequations}

\section{Explicit formulas for spin-3 field}\label{app:C}

The non-zero components of the spin-3 field in \eqref{spin3} are given by
\begin{subequations}
  \begin{align}
    \Phi_{vvv} &= 54 r \mu_\mathcal{J} \mu^{(3)}_{\mathcal{J}} \mu_\mathcal{P} +9\mu_\mathcal{J}^2 \mu_\mathcal{P} \mu^{(3)}_{\mathcal{P}}-36 r (\mu^{(3)}_{\mathcal{J}})^2 \mu^{(3)}_{\mathcal{P}}-4 \mu _P (\mu^{(3)}_{\mathcal{P}})^3-9 r \mu_\mathcal{P}^2 \mu^{(3)}_{\mathcal{P}} \\
    \Phi_{vv\varphi} &= -72 \mathcal{J}_3 r \mu^{(3)}_{\mathcal{J}} \mu^{(3)}_{\mathcal{P}} +54 \mathcal{J} r \mu_\mathcal{J} \mu^{(3)}_{\mathcal{J}}+9 \mathcal{P}^{(3)} \mu_{\mathcal{J}}^2 \mu_\mathcal{P} +18 \mathcal{P} \mu_\mathcal{J} \mu _\mathcal{P} \mu^{(3)}_{\mathcal{P}}-18 \mathcal{J} r \mu_\mathcal{P} \mu^{(3)}_{\mathcal{P}}  \nonumber\\
                &\;\, +54 \mathcal{J}^{(3)} r \mu_\mathcal{J} \mu _\mathcal{P} -36 \mathcal{P}^{(3)} r (\mu^{(3)}_{\mathcal{J}})^2+54 \mathcal{P} r \mu^{(3)}_{\mathcal{J}} \mu_\mathcal{P} -12 \mathcal{P}^{(3)} \mu_\mathcal{P} (\mu^{(3)}_{\mathcal{P}})^2 -9 \mathcal{P}^{(3)} r \mu_\mathcal{P}^2 \\
    \Phi_{vvr} &= 27 \mu_\mathcal{J} \mu^{(3)}_{\mathcal{J}}-9 \mu _P \mu^{(3)} _{\mathcal{P}} \displaybreak[1]\\
    \Phi_{v\varphi\varphi} &= -9 \mathcal{J}^2 r \mu^{(3)}_{\mathcal{P}}+54 \mathcal{J} \mathcal{P} r \mu^{(3)}_{\mathcal{J}}-72 \mathcal{J}^{(3)}\mathcal{P}^{(3)} r \mu^{(3)}_{\mathcal{J}}+18 \mathcal{P} \mathcal{P}^{(3)} \mu_\mathcal{J} \mu_\mathcal{P} -18\mathcal{J} \mathcal{P}^{(3)} r \mu_\mathcal{P}   \nonumber \\
                &\quad +54 \mathcal{J}^{(3)} \mathcal{P} r \mu_\mathcal{P} -36 (\mathcal{J}^{(3)})^2 r \mu^{(3)}_{\mathcal{P}}+54 \mathcal{J}^{(3)} \mathcal{J} r \mu_\mathcal{J} +9 \mathcal{P}^2 \mu _\mathcal{P} \mu^{(3)}_{\mathcal{P}}-12 {(\mathcal{P}^{(3)})}^2 \mu_\mathcal{P} \mu^{(3)}_{\mathcal{P}} \displaybreak[1]\\
    \Phi_{v\varphi r} &= 27 {\mathcal{J}}^{(3)} \mu_{\mathcal{J}} -9 \mathcal{J} \mu^{(3)}_{\mathcal{P}}+27 \mathcal{P} \mu^{(3)}_{\mathcal{J}}-9 \mathcal{P}^{(3)} \mu_\mathcal{P} \displaybreak[1]\\
    \Phi_{\varphi\varphi\varphi} &= -9 \mathcal{J}^2 \mathcal{P}^{(3)} r -36 (\mathcal{J}^{(3)})^2 \mathcal{P}^{(3)} r+54 \mathcal{J} \mathcal{J}^{(3)} \mathcal{P} r+9 \mathcal{P}^2 \mathcal{P}^{(3)} \mu_\mathcal{P} -4 (\mathcal{P}^{(3)})^{3} \mu_\mathcal{P} \displaybreak[1]\\
    \Phi_{\varphi\varphi r} &= 27 \mathcal{J}^{(3)}\mathcal{P} -9 \mathcal{J} \mathcal{P}^{(3)}\,.
  \end{align}
\end{subequations}

\bibliographystyle{fullsort}
\bibliography{Bibliography}

\end{document}